\documentclass[10pt,journal,final,twocolumn]{IEEEtran}

\usepackage[T1]{fontenc}
\usepackage{graphicx}
\usepackage{amsfonts,amssymb,amsmath}
\usepackage{amsthm}
\usepackage{mathtools}
\usepackage{subcaption}
\usepackage{setspace}
\usepackage{color}
\usepackage{algorithm, algorithmic, tabularx}
\usepackage{bm}
\usepackage{cite}

\makeatletter
\def\endthebibliography{%
  \def\@noitemerr{\@latex@warning{Empty `thebibliography' environment}}%
  \endlist
}
\makeatother

\begin{document}
\title{\huge IRS-Assisted Ambient Backscatter Communications Utilizing Deep Reinforcement Learning}

\author{Xiaolun Jia,~\IEEEmembership{Graduate Student Member,~IEEE,}
        and~Xiangyun Zhou,~\IEEEmembership{Senior Member,~IEEE}
        \vspace{-6mm}
\thanks{The authors are with the School of Engineering, The Australian National University, Canberra, Australia (e-mail: xiaolun.jia@anu.edu.au).}
}

\maketitle

\begin{abstract}
We consider an ambient backscatter communication (AmBC) system aided by an intelligent reflecting surface (IRS). The optimization of the IRS to assist AmBC is extremely difficult when there is no prior channel knowledge, for which no design solutions are currently available. We utilize a deep reinforcement learning-based framework to jointly optimize the IRS and reader beamforming, with no knowledge of the channels or ambient signal. We show that the proposed framework can facilitate effective AmBC communication with a detection performance comparable to several benchmarks under full channel knowledge.
\end{abstract}


\IEEEpeerreviewmaketitle

\vspace{-2mm}
\section{Introduction}

Ambient backscatter communication (AmBC) is a key enabler for energy-efficient networking in the Internet of Things, where AmBC devices (or tags) convey information on top of modulated radiofrequency (RF) signals \cite{Yang18}. Despite this, a key problem is that the backscattered information signal at the reader experiences severe direct-link interference (DLI) from the ambient signal, which may be unknown and orders of magnitude stronger, resulting in poor detection performance.

Various methods, from transceiver design to channel estimation and modifications to network infrastructure, have been proposed to improve the AmBC detection performance. Work in \cite{Hes19} proposed a tag which performs frequency shift modulation to separate the backscattered signal from the DLI, at the cost of tag complexity. Machine learning-inspired approaches were presented in \cite{Guo19, blind}. Recently, the use of an intelligent reflecting surface (IRS) to improve AmBC performance was also explored in \cite{Zhao20, Nem20}.

IRSs have received significant research interest due to their ability to impose variable phase shifts on impinging signals to obtain desired reception at a receiver \cite{hmimos, Huang19}. The joint phase shift optimization of a large number of reflectors allow favorable received signal strength scaling, proportional to the IRS area \cite{Ozd19}. A further advantage of reconfigurable phase shifts is the ability to focus signals in different directions to reduce inter-user interference \cite{Wu19a}. This makes IRS an ideal candidate to address the problem of severe DLI in AmBC systems, and hence improve AmBC detection.

In this letter, we study the signaling design and phase shift optimization of an IRS-assisted AmBC system in indoor settings \cite{indoor} such as smart homes, to improve the detection performance. A passive IRS is used for energy efficiency. Typically, IRS optimization requires full channel state information (CSI). However, the AmBC component poses severe challenges to CSI acquisition: a) channels between all nodes exist in typical AmBC systems, but the channels involving the ambient source are often unknown and its signal varies every sample; b) as the tag is also a passive reflector, it is difficult to resolve the IRS-tag channels. For these reasons, many channels are very hard to estimate, and thus the CSI is assumed to be unavailable, rendering IRS optimization a formidable challenge. Our work presents, to our best knowledge, the first solution to optimizing the IRS under these realistic AmBC conditions. Work in \cite{Chen21} is the only other study on IRS optimization in AmBC systems, but nonetheless adopted a full CSI assumption; while work in \cite{Guo19} optimized the AmBC detection without CSI, but in a system with no IRS. We perform the optimization with a deep reinforcement learning (DRL) approach using the deep deterministic policy gradient (DDPG) algorithm. DRL can solve highly complex problems involving IRSs such as the joint optimization of phase shifts and other parameters \cite{Huang20}, and more generally, problems with large numbers of control variables and only partially observable environments \cite{Lil15}. We utilize AmBC domain knowledge to propose a modified DDPG algorithm compared to those in conventional IRS works with known CSI \cite{Huang20, Feng20}, which works off individual signal samples. Our results show that the proposed approach performs comparably to several full-CSI benchmarks.

\vspace{-2mm}
\section{System Model and Problem Formulation}

\begin{figure}[!h]
\centering
\includegraphics[width=3.5in]{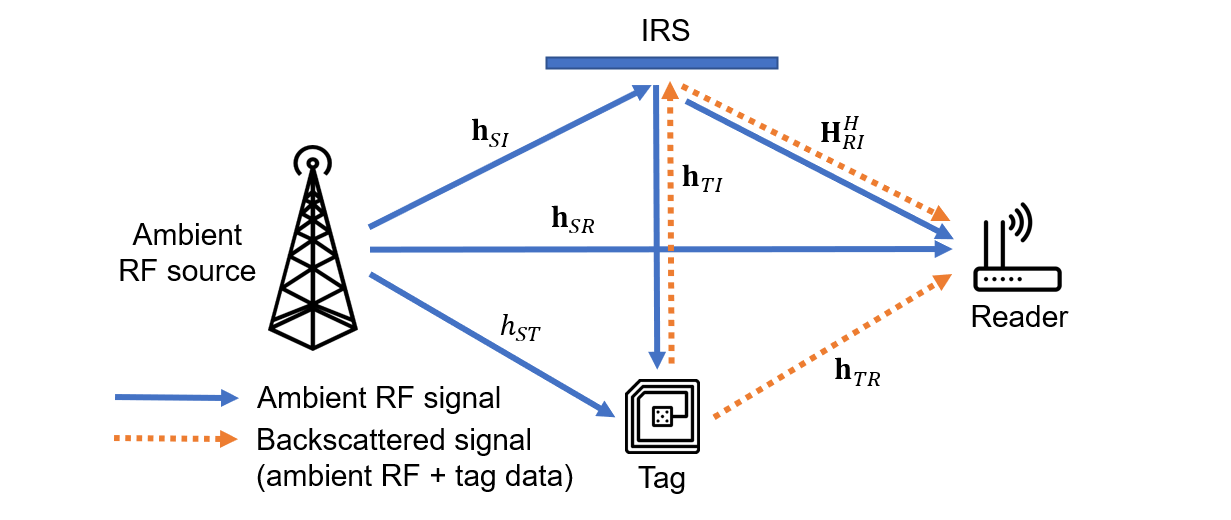}
\caption{IRS-assisted AmBC system.}
\label{fig:system_setup}
\end{figure}

Consider an AmBC system in Fig.~\ref{fig:system_setup} with an ambient RF source, a single-antenna tag, an IRS with $N$ reflectors and a reader with $M$ antennas. Hereafter, we assign subscripts $S$, $T$, $I$ and $R$ to the source, tag, IRS and reader, respectively.

We adopt a discrete-time signal model, as in \cite{Qian17a}. The source transmits signal $s[\ell]$, whose samples are i.i.d. and follow $\mathcal{CN}(0, P_{S})$, where $P_{S}$ is the transmit power. The assumption of a complex Gaussian ambient signal is commonly adopted in AmBC systems \cite{Wang16, Qian17a, Guo19}. 

The tag is equipped with two impedances and performs on-off keying modulation, with the data symbols being either $0$ or $1$. For ease of exposition, we assume that the tag has a built-in battery, which powers the circuit operation over a typical lifetime of several years. As the tag is a diffuse scatterer, we take the strengths of signal paths undergoing two or more reflections at the tag before reaching the reader to be negligible. However, this assumption does not apply at the IRS, due to its ability to enhance the overall signal strength while balancing between different reflection links.

Linear combining is performed at the reader using the vector $\mathbf{g} \in \mathbb{C}^{M \times 1}$, where we set $\left\lVert \mathbf{g} \right\rVert^{2} = 1$ without loss of generality. Conventional energy detection \cite{Guo19} is utilized to recover the backscattered data symbols after applying the combiner.

All channels are assumed to undergo frequency-flat quasi-static fading. The channels from the source to tag, source to IRS, source to reader, tag to reader, tag to IRS and IRS to reader are $h_{ST} \in \mathbb{C}^{1 \times 1}$, $\mathbf{h}_{SI} \in \mathbb{C}^{N \times 1}$, $\mathbf{h}_{SR} \in \mathbb{C}^{M \times 1}$, $\mathbf{h}_{TR} \in \mathbb{C}^{M \times 1}$, $\mathbf{h}_{TI} \in \mathbb{C}^{N \times 1}$ and $\mathbf{h}_{RI}^{H} \in \mathbb{C}^{M \times N}$, respectively. Moreover, we assume that the ambient source does not cooperate with the system, and that the reader has no knowledge of the ambient signal or the CSI of any channel in the system.

Each IRS reflector has a reconfigurable phase shift, denoted by $\theta_{n} \in [0, 2\pi)$ for the $n$-th reflector. The incident signal at each reflector is subject to the reflection coefficient $\Theta_{n} = e^{j \theta_{n}}$ arising from the phase shift. The amplitude scaling of all IRS reflectors is set to unity. Thus, the matrix of reflection coefficients at the IRS is $\mathbf{\Theta} = \mathrm{diag}(\Theta_{1}, \ldots, \Theta_{N})$.

The signal received by the tag, consisting of the direct source-tag and reflected source-IRS-tag signal paths, is
\begin{equation}
y_{T}[\ell] = \left( \mathbf{h}_{TI}^{H} \mathbf{\Theta} \mathbf{h}_{SI} + h_{ST} \right) s[\ell]. 	\label{tagReceived}
\end{equation}
The tag backscatters $x_{T}[\ell] = \alpha b[\ell] y_{T}[\ell]$, where $\alpha$ denotes the tag splitting coefficient, which is set to $1$ without loss of generality; and $b[\ell]$ denotes the $\ell$-th sample of the backscattered data symbol. We assume that the duration of one backscatter symbol spans $L$ samples. Therefore, denoting the $k$-th backscatter symbol as $b_{k}$, we have $b[\ell] = b_{k} \in \{0, 1\}, \forall \ell \in \{(k-1)L + 1, \ldots, kL\}$. 

The reader receives the ambient RF signal from the source-reader and source-IRS-reader paths, plus the backscattered signals from four paths, as combinations of \{source-tag, source-IRS-tag\} multiplied by \{tag-reader, tag-IRS-reader\} paths:
\begin{equation}
\mathbf{y}_{R}[\ell] = 
	\begin{dcases}
		\mathbf{h}_{0} s[\ell] + \mathbf{z}_{R}[\ell], & b[\ell] = 0, \\
		(\mathbf{h}_{0} + \alpha \mathbf{h}_{1}) s[\ell] + \mathbf{z}_{R}[\ell], & b[\ell] = 1,
	\end{dcases}	\label{receivedBits}
\end{equation}
with $\mathbf{h}_{0} = \mathbf{H}_{RI}^{H} \mathbf{\Theta} \mathbf{h}_{SI} + \mathbf{h}_{SR}$ and $\mathbf{h}_{1} = \left( \mathbf{H}_{RI}^{H} \mathbf{\Theta} \mathbf{h}_{TI} + \mathbf{h}_{TR} \right) \times (\mathbf{h}_{TI}^{H} \mathbf{\Theta} \mathbf{h}_{SI} + h_{ST})$, where $\mathbf{z}_{R}[\ell] \sim \mathcal{CN}(0, P_{w} \mathbf{I})$ is the noise at the reader. The final signal after the combiner is given by $\mathbf{g}^{H} \mathbf{y}_{R}$. We adopt the shorthand $\mathbf{h}_{A} \triangleq \mathbf{h}_{0}$, $\mathbf{h}_{I} \triangleq \alpha \mathbf{h}_{1}$ and $\mathbf{h}_{AI} \triangleq \mathbf{h}_{A} + \mathbf{h}_{I}$, where subscripts $A$ and $I$ represent the \underline{a}mbient and \underline{i}nformation-bearing components, respectively.

As per conventional AmBC systems (e.g., \cite{Wang16}), the energies of the received $0$ and $1$ symbols are modeled as Gaussian random variables, with means and variances given by 
\begin{equation}
\mu_{0} = P_{s} |\mathbf{g}^{H} \mathbf{h}_{A}|^{2} + P_{w},	\hspace{5mm} \sigma_{0}^{2} = \mu_{0}^{2} / L,		\label{stats0}
\end{equation}
\begin{equation}
\mu_{1} = P_{s} |\mathbf{g}^{H} (\mathbf{h}_{A} + \alpha \mathbf{h}_{I})|^{2} + P_{w}, 	\hspace{5mm} \sigma_{1}^{2} = \mu_{1}^{2} / L.		\label{stats1}
\end{equation}
We define the \textit{generalized relative channel difference (GRCD)} as $\Delta_{G} = \max \big\{ \frac{\mu_{1}}{\mu_{0}}, \frac{\mu_{0}}{\mu_{1}} \big\}$, which is the energy ratio between the symbol with the higher energy and the symbol with the lower energy. We note that the GRCD directly determines the BER of the AmBC system. Under a central limit theorem assumption and reasonably large $L$, the BER is derived from \cite{Guo19} as
\begin{equation}
p_{b}\!=\!\frac{1}{2}\! \Big[\!Q_\mathcal{N} \Big(\!\sqrt{L} \Big( \frac{\Delta_{G} \log \Delta_{G}}{\Delta_{G} - 1}\!-\!1 \Big)\!\Big)\!+\!Q_\mathcal{N} \Big(\!\sqrt{L} \Big( 1\!-\!\frac{\log \Delta_{G}}{\Delta_{G} - 1}  \Big)\!\Big)\!\Big],	\label{BER}
\end{equation}
where $Q_\mathcal{N}(\cdot)$ is the Gaussian $Q$-function. One can show that the BER reduces as GRCD increases. Hence, we aim to maximize the GRCD of the IRS-assisted AmBC system (which minimizes the BER) by jointly designing the IRS and reader parameters in the following problem:
\begin{subequations}
\begin{align}
\text{(P)}: ~~\max_{\mathbf{g}, \mathbf{\Theta}} ~~~&\Delta_{G} \label{PA}\\
\mathrm{s.t.}~~~~& |\Theta_{n}| = 1, \forall n \in \{1, \ldots, N\},  \label{PB} \\
&\!\left\lVert \mathbf{g} \right\rVert^{2} = 1. 	\label{PC}
\end{align}
\end{subequations}
Note that $\Delta_{G}$ in Problem (P) can only be obtained when the full CSI is available. When no CSI is available, we use instantaneous signal observations to estimate the ``sample'' GRCD for the $t$-th time step, denoted by $\Delta_{G}^{(t)}$. The sample GRCD is estimated using $\mu_{0}$ and $\mu_{1}$ averaged over $L$ samples in each symbol. As our work assumes the absence of CSI, we propose a DRL-based solution in the next section, by approximating $\Delta_{G}$ based on instantaneous signal observations.

\vspace{-2mm}
\section{DRL-Based Framework}

\subsection{Reinforcement Learning Fundamentals}

Reinforcement learning problems concern the interactions between an agent and the environment in order to maximize a reward, which can be formulated as a Markov decision process (MDP). At time step $t$, the state $s_{t}$ characterizes the environment. Based on $s_{t}$, the agent takes an action $a_{t}$ drawn from a policy $\pi$. The action influences the environment, which takes on a new state $s_{t+1}$; while a reward $r_{t}$ is provided to the agent. The agent stores experiences over time, each of the form $\langle s_{t}, a_{t}, r_{t}, s_{t+1}\rangle$. The agent aims to determine an optimal policy $\pi_{*}: \mathcal{S}\!\rightarrow\!\mathcal{A}$ to maximize the action-value function and thus the expected discounted reward (return), given by
\begin{equation}
Q_{\pi}(s_{t}, a_{t}) = \mathbb{E}\left\{{\textstyle\sum}_{k=0}^{\infty} \gamma^{k} r_{t+k+1} | s_{t} = s, a_{t} = a\right\},	\label{qfunc}
\end{equation}
where $\gamma \in [0, 1]$ denotes the discount factor.

\begin{figure}[!h]
\centering
\includegraphics[width=3.5in]{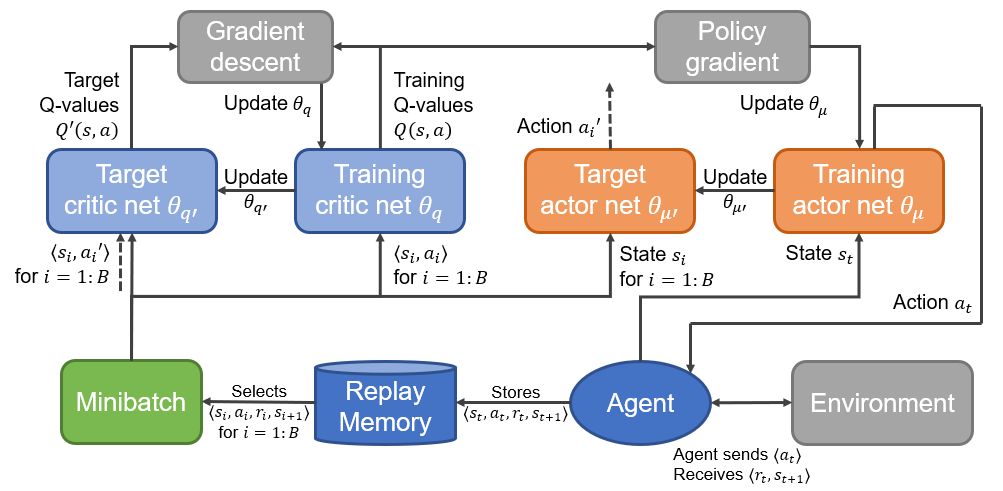}
\caption{Diagram of the DDPG algorithm.}
\label{fig:ddpg}
\end{figure}

DDPG is an algorithm applicable to MDPs with continuous action spaces \cite{Lil15}. Two components, the actor and critic, simultaneously learn the policy and $Q$-functions, respectively. Each consists of two deep neural networks (DNNs), termed the training and target nets, as shown in Fig.~\ref{fig:ddpg}.

Denote the parameters (weights) of the training and target actor nets by $\theta_{\mu}$ and $\theta_{\mu'}$, respectively, and those of the training and target critic nets by $\theta_{q}$ and $\theta_{q'}$, respectively. Here, we use the notation $\mu(s_{t})$ instead of $\pi$ to denote the policy, to highlight the continuous action space. The agent stores past experiences in its replay memory $\mathcal{E}$. At each time step, the agent samples a minibatch of $B$ experiences from $\mathcal{E}$ and computes target $Q$-values for each experience using
\begin{equation}
y_{i} = r_{i} + \gamma Q'(s_{i+1}, \mu'(s_{i+1} | \theta_{\mu'}) | \theta_{q'}),	\label{targetQ}
\end{equation}
where $Q'(s, a)$ is the $Q$-value from the target critic net. Next, gradient descent is performed on $\theta_{q}$ 
to minimize the overall loss between the target $Q$-values and those produced by the training net, with the loss function given by
\begin{equation}
\mathcal{L}(\theta_{q}) = (1/B) {\textstyle\sum}_{i=1}^{B} \left( y_{i} - Q(s_{i}, a_{i} | \theta_{q}) \right)^{2}.	\label{criticLoss}
\end{equation}
Subsequently, the training actor net, which produces the agent's policy, is updated by sampling the policy gradient (PG), which provides an approximation of the policy's expected return. Maximizing the return involves performing gradient ascent, whose update rule is given by
\begin{equation}
\theta_{k+1} = \theta_{k} + \alpha_{\mu} \nabla_{a} Q(s_{t}, \mu(s_{t} | \theta_{\mu}) | \theta_{q}) \nabla_{\theta_{\mu}} \mu(s_{t} | \theta_{\mu}),	\label{actorPGUpdate}
\end{equation}
where $\alpha_{\mu}$ is the learning rate and the subsequent terms approximate the PG. Finally, to ensure stability during training, the target nets are updated every $T_{up}$ time steps according to
\begin{equation}
\theta_{a'} = \tau \theta_{a} + (1 - \tau) \theta_{a'}, \hspace{5mm} a \in \{\mu, q\},	\label{SoftUpdate}
\end{equation}
where $\tau \ll 1$ is the update coefficient for the target nets. 

\vspace{-3mm}
\subsection{DRL Problem Reformulation and Proposed Algorithm}

Problem (P) can be formulated as an MDP, where the reader acts as the agent and is responsible for the joint design of the IRS reflection coefficients and its own combiner as its action. The reader runs the proposed DRL algorithm to achieve this design by interacting with the wireless propagation environment, which is characterized by the channels.

Each channel coherence period, consisting of $T$ time steps, is defined as one episode. Due to the stochastic and unknown nature of the channels, each episode is independent from others, as the underlying environment varies with the channels in each episode. Thus, a key difference of the DRL framework in our work compared to existing DRL works involving IRS (e.g., \cite{Huang20, Tah20, Feng20}), where the CSI is known, is the fact that the actor and critic are trained in each episode based only on the observations in the current episode.
\begin{itemize}
\item \textit{State:} The current state is the concatenation of the previous combiner and IRS reflection coefficients ($2M + 2N$ elements). As existing DNN implementations do not support complex inputs, the real and imaginary components of combiner weights and reflection coefficients are used:
\begin{multline}
s_{t} = [\mathrm{Re}\{[g_{1}^{(t-1)}, \ldots, g_{M}^{(t-1)}, \Theta_{1}^{(t-1)}, \ldots, \Theta_{N}^{(t-1)}]\}, \\
 \mathrm{Im}\{[g_{1}^{(t-1)}, \ldots, g_{M}^{(t-1)}, \Theta_{1}^{(t-1)}, \ldots, \Theta_{N}^{(t-1)}]\}].  \label{state}
\end{multline}

\item \textit{Action}: The action space consists of the real and imaginary components of only the updated reflection coefficients ($2N$ elements) based on the current state, for reasons which will be explained in the sequel. That is,
\begin{equation}
a_{t} = [ \mathrm{Re}\{[\Theta_{1}^{(t)}, \ldots, \Theta_{N}^{(t)}]\}, \mathrm{Im}\{[\Theta_{1}^{(t)}, \ldots, \Theta_{N}^{(t)}]\} ].		\label{action}
\end{equation}

\item \textit{Reward}: Instead of using $\Delta_{G}^{(t)}$ directly as the reward, we modify the reward function to $r_{t} = 100 (\Delta_{G}^{(t)} - 1)$. From our experimentation, we observed that when random combiner weights and reflection coefficients were used, the majority of GRCD values were close to $1$. This may lead to underfitting when different states and actions result in similar rewards. The multiplicative factor of $100$ thus spreads out the reward space, such that each reward may be relatively more distinct, allowing faster convergence to be achieved by the critic nets.
\end{itemize}

In our experimentation with various state and action spaces, we observed poor performance when the combiner and reflection coefficients were set together. Thus, inspired by \cite{Guo19}, we propose to pre-set the combiner to the optimal eigenvector beamformer corresponding to the signal observations in the current DRL step, independent from the reflection coefficients (which are updated later). This has two practical advantages. First, the optimal combiner results in a relatively large GRCD (compared to e.g., a random combiner), which can be further improved through tuning the reflection coefficients. Second, the $Q$-function, which relies on both state and action, is derived from only one combiner, as opposed to one from the current state and one from the action (which may be vastly different), resulting in more effective fitting of the $Q$-function.

The $T$ time steps in each episode are divided into training and data transmission phases. In the training phase, each step is a DRL agent-environment interaction, and consists of two pairs of backscatter pilot symbols, with each pair being a $0$ followed by a $1$, and each symbol spanning $L_{t}$ samples. The estimated channel covariance matrices for the first pilot pair, denoted by $\mathbf{C}_{i}, \ i \in \{0, 1\}$, are first obtained using the current reflection coefficients (i.e., $\mathbf{\Theta}^{(t-1)}$), and given by
\begin{equation}
\mathbf{C}_{i} = (1/L_{t}) {\textstyle\sum}_{\ell = 1}^{L_{t}} \mathbf{y}_{R}[\ell] \mathbf{y}_{R}[\ell]^{H} |_{b_{k} = i}, 	\label{sampleCovariance}
\end{equation}
and are then refined using the maximum eigenvalue and corresponding eigenvector \cite{Guo19}. Then, we update the combiner by solving the equation $\mathbf{C}_{1} \mathbf{g} = \lambda \mathbf{C}_{0} \mathbf{g}$, where $\lambda$ represents the generalized eigenvalues of $\{\mathbf{C}_{0},  \mathbf{C}_{1}\}$. The combiner is set to the eigenvector of the maximum eigenvalue $\lambda^{+}$ if $\lambda^{+} > 1 / \lambda^{-}$, with $\lambda^{-}$ being the minimum eigenvalue, and the eigenvector for $\lambda^{-}$ otherwise. The updated combiner is concatenated with the current IRS reflection coefficients to give an `intermediate' state $s_{t,\mathrm{int}}$. We then feed this intermediate state into the actor to obtain the updated reflection coefficients, which is the action for this DRL step. The second pair of pilot symbols is then transmitted, yielding $\mathbf{C}_{i}^{'}, \ i \in \{0, 1\}$, and refined similarly. The sample GRCD for the current DRL step is then obtained by evaluating $\Delta_{G}^{(t)} = \max \Big\{ \frac{\mathbf{g}^{H} \mathbf{C}_{1}^{'} \mathbf{g}}{\mathbf{g}^{H} \mathbf{C}_{0}^{'} \mathbf{g}}, \frac{\mathbf{g}^{H} \mathbf{C}_{0}^{'} \mathbf{g}}{\mathbf{g}^{H} \mathbf{C}_{1}^{'} \mathbf{g}} \Big\}$.

As the underlying channels are unknown, we reserve the first $T_{train}$ time steps of each episode, where the agent explores while training the actor and critic to learn the current reward function. The $T_{train}$ steps are divided into two phases: the agent takes random actions for the first $T_{1}$ steps, followed by actions generated from the actor for $T_{2}$ steps. After $T_{train}$ steps, the final set of $\{\mathbf{g}, \mathbf{\Theta}\}$ is fixed for the remainder of the episode, which is the data transmission phase, where a shorter symbol duration of $L_{d}$ samples is used to provide a higher data rate. We note that the use of a random training phase is critical to achieve desirable results for this problem.

The DRL approach is presented in Algorithm 1. In each episode, the reader begins by instructing the tag to transmit a fixed number of pilot symbol pairs, followed by its message. Once initiated, the tag only needs to transmit its pilots and data, without further interactions with the system. A control link exists between the reader and the IRS for sending the phase shift instructions during each DRL step. The DDPG algorithm complexity is $O(2 T_{train} B ({\textstyle\sum}_{y=1}^{Y-1} u_{y} u_{y+1}))$ per DNN per episode from forward and backward propagation, where $u_{y}$ is the number of hidden units in layer $y$. The computation of the covariance matrices incurs very small cost compared to forward and backward propagation; as such, the complexity per episode is similar to the DDPG variants in e.g., \cite{Huang20}.

{\singlespacing
\begin{algorithm}
	\caption{DDPG Algorithm for IRS-AmBC Design}
	\begin{algorithmic}[1]
		\STATE \textbf{Inputs:} Replay memory, $\mathcal{E}$; minibatch size, $B$; actor and critic learning rates, $\alpha_{\mu}$ and $\alpha_{q}$; update coefficient for target nets, $\tau$; discount factor, $\gamma$; noise process, $\mathcal{N}$
		\FOR{each episode}
			\STATE Initialize training nets $\mu(s | \theta_{\mu})$, $Q(s, a | \theta_{q})$; target nets $\mu'(s | \theta_{\mu'}) = \mu(s | \theta_{\mu})$, $Q'(s, a | \theta_{q'}) = Q(s, a | \theta_{q})$; random initial values $\mathbf{g}^{(0)}$ and $\mathbf{\Theta}^{(0)}$; empty replay memory $\mathcal{E}$
			\FOR{time step $t = 1$:$T_{train}$}
				\STATE Observe $\mathbf{C}_{0}$ and $\mathbf{C}_{1}$ and set $\mathbf{g}^{(t)}$ to the optimal eigenvector beamformer
				\STATE Set intermediate state $s_{t,\mathrm{int}} \leftarrow \{\mathbf{g}^{(t)}, \mathbf{\Theta}^{(t-1)}\}$
				\STATE Observe $\mathbf{C}_{0}^{'}$ and $\mathbf{C}_{1}^{'}$. \textbf{If} $t < T_{1}$ then take random action; \textbf{else} take action $a_{t} = \mu(s_{t,i} | \theta_{\mu}) + \mathcal{N}$
				\STATE Set next state $s_{t+1} \leftarrow \{\mathbf{g}^{(t)}, a_{t}\}$ and store experience $\langle s_{t}, a_{t}, r_{t}, s_{t+1} \rangle$ in $\mathcal{E}$
				\STATE Sample a minibatch $\mathcal{B}$ of $B$ experiences from $\mathcal{E}$
				\STATE Set target $Q$-values for minibatch $\mathcal{B}$ according to (\ref{targetQ})
				\STATE Perform gradient descent on training critic net by minimizing loss function in (\ref{criticLoss})
				\STATE Update training actor net using sampled PG in (\ref{actorPGUpdate})
				\STATE Update target actor and critic nets according to (\ref{SoftUpdate})
				\STATE Update state $s_{t} \leftarrow s_{t+1}$
			\ENDFOR
		\ENDFOR
	\end{algorithmic}
\end{algorithm}
}


\vspace{-3mm}
\section{Numerical Results}

We demonstrate the performance of the DRL framework with the following simulation setup: all IRS channels undergo Rician fading as in \cite{Chen21} with Rician factor $3$; while all other channels experience Rayleigh fading, typical of scattering occurring in a smart home environment. The path loss exponent is $2.5$ for all channels. The ambient signal frequency is $2.4$ GHz with transmit power $P_s = 20$ dBm, typical of a Wi-Fi access point. The source, tag, IRS and reader are located at $[-5, 0]$, $[0, 0]$, $[0, 5]$ and $[5, 0]$ m, respectively. The reader has $M = 4$ antennas and the noise variance is $P_w = -95$ dBm. We group sub-groups of IRS reflectors to take on the same reflection coefficients for efficient computation \cite{Yang20}, such that each of the $N$ effective reflectors is one-wavelength-sized.

The training and target nets in the actor and critic are fully-connected DNNs, with an input layer, two hidden layers and an output layer. The sizes of the actor and critic nets are $[2M\!+\!2N, 4M\!+\!4N, 4M\!+\!4N, 2N]$ and $[2M\!+\!4N, 4M\!+\!8N, 4M\!+\!8N, 1]$, respectively. The hidden and output layers use the rectified linear unit (ReLU) and linear activation functions, respectively. After linear activation, each pair of outputs from the actor nets, corresponding to one reflection coefficient, is $\ell_{2}$-normalized to unit magnitude. The DNN parameters are $\alpha_{\mu}, \alpha_{q}\!=\!0.002$, $\tau\!=\!0.0005$, $T_{up}\!=\!1$, $B\!=\!16$; and the optimizer is RMSprop with momentum $0.8$. The policy noise process $\mathcal{N}$ is the Ornstein-Uhlenbeck process with standard deviation $0.05$. $1000$ channel realizations are used, with $\{ T_{1}, T_{2} \}\!=\!\{ 1000, 500 \}$ steps. Unique to our problem, under unknown CSI, the reward function and its maximum attainable reward vary with each channel realization. In order for the critic nets to properly fit the reward function for each channel realization, the $Q$-values must be based on the instantaneous reward, resulting in $\gamma = 0$ \cite{Tah20}. 

\begin{figure}[!h]
\centering
\includegraphics[width=3.5in]{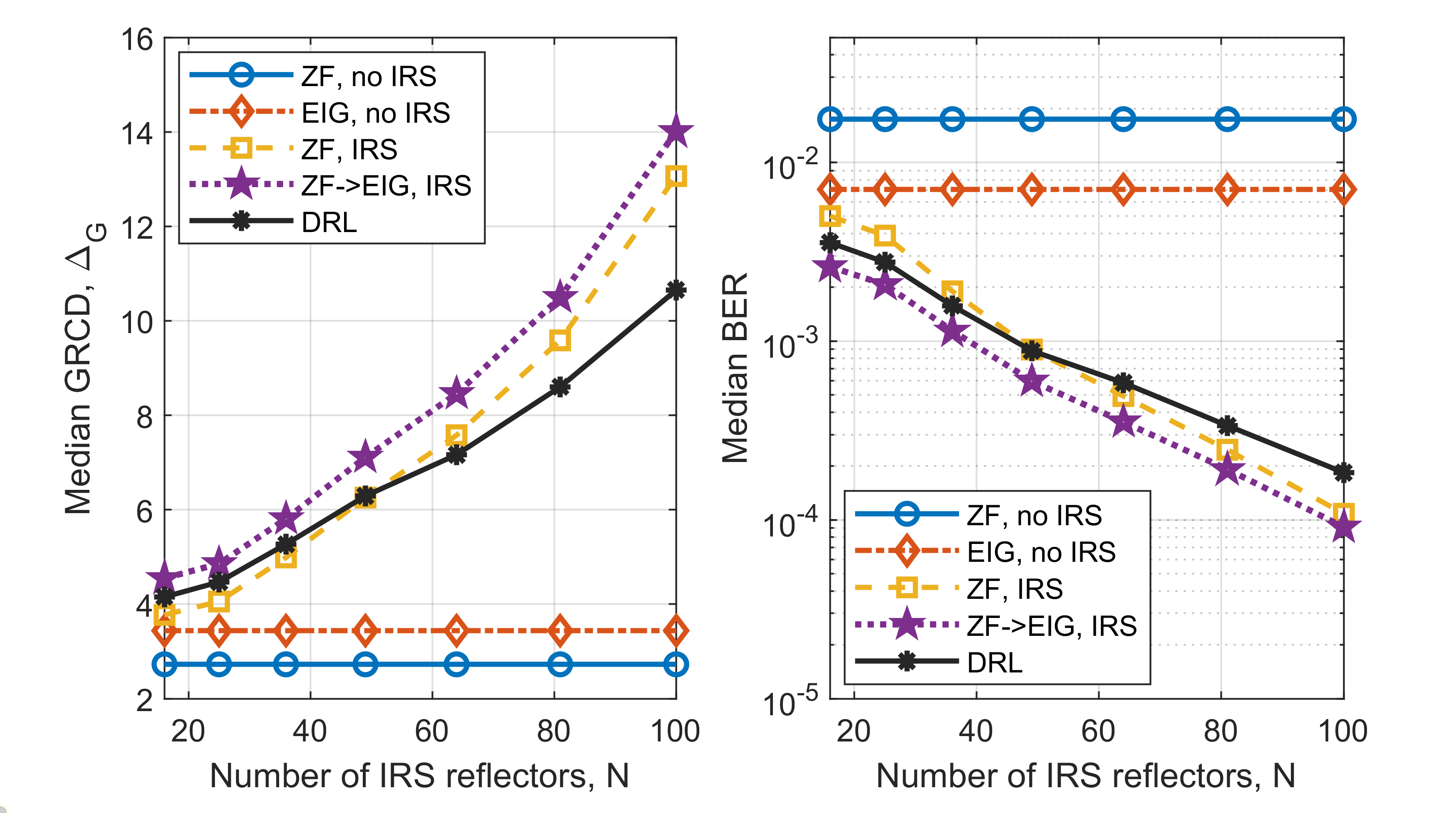}
\caption{(a) Median GRCD vs. $N$; (b) Median BER vs. $N$.}
\label{fig:3}
\end{figure}

Fig.~\ref{fig:3} highlights the effect of the number of IRS reflectors on the GRCD and BER in the training and data transmission phases, respectively. For this experiment, random samples of $s[\ell]$ and $\mathbf{z}_{R}[\ell]$ are generated in each backscatter symbol period. The symbol durations for the training and data transmission phases are $\{ L_{t}, L_{d} \} = \{ 150, 20 \}$, with a larger $L_{t}$ needed to accurately estimate the channel covariance matrices in (\ref{sampleCovariance}). With our values of $\{ T_{1},\!T_{2} \}$, the training phase is $<\!20\%$ of a typical channel coherence time of $100$ ms \cite{Alev17} for backscatter setups with stationary nodes and a software-defined radio-type reader. In addition to the results from Algorithm 1, four benchmarks are included for comparison. These are: 1) the optimal zero-forcing (ZF) and 2) eigenvector (EIG) combiners when no IRS is present, 3) the ZF combiner with IRS, and 4) the EIG combiner with IRS, initialized using the $\mathbf{g}$ and $\mathbf{\Theta}$ solutions from 3). All four benchmarks are obtained under full CSI and average noise power. Due to the large variance in GRCD values as a result of small-scale fading, the median GRCD is shown for each IRS size and benchmark. One should note that with no CSI on top of the varying ambient signal and noise, the DRL framework can never outperform the full-CSI benchmarks. However, Fig.~\ref{fig:3}(a) shows that the proposed CSI-free DRL framework still performs within $25\%$ of the best benchmark under full CSI (i.e., Benchmark 4) for the range of $N$ values. This is a significant result that illustrates the effectiveness of the proposed CSI-free framework.

Fig.~\ref{fig:3}(b) translates the median GRCD into the equivalent BER in the data transmission phase using (\ref{BER}). We find that the BER achieved with Algorithm 1 is comparable with the best benchmark for all values of $N$. Moreover, one order-of-magnitude BER improvement may be achieved using a moderately-sized IRS with $N = 64$, which is a significant gain over the best non-IRS benchmark.

\begin{figure}[!h]
\centering
\includegraphics[width=3.5in]{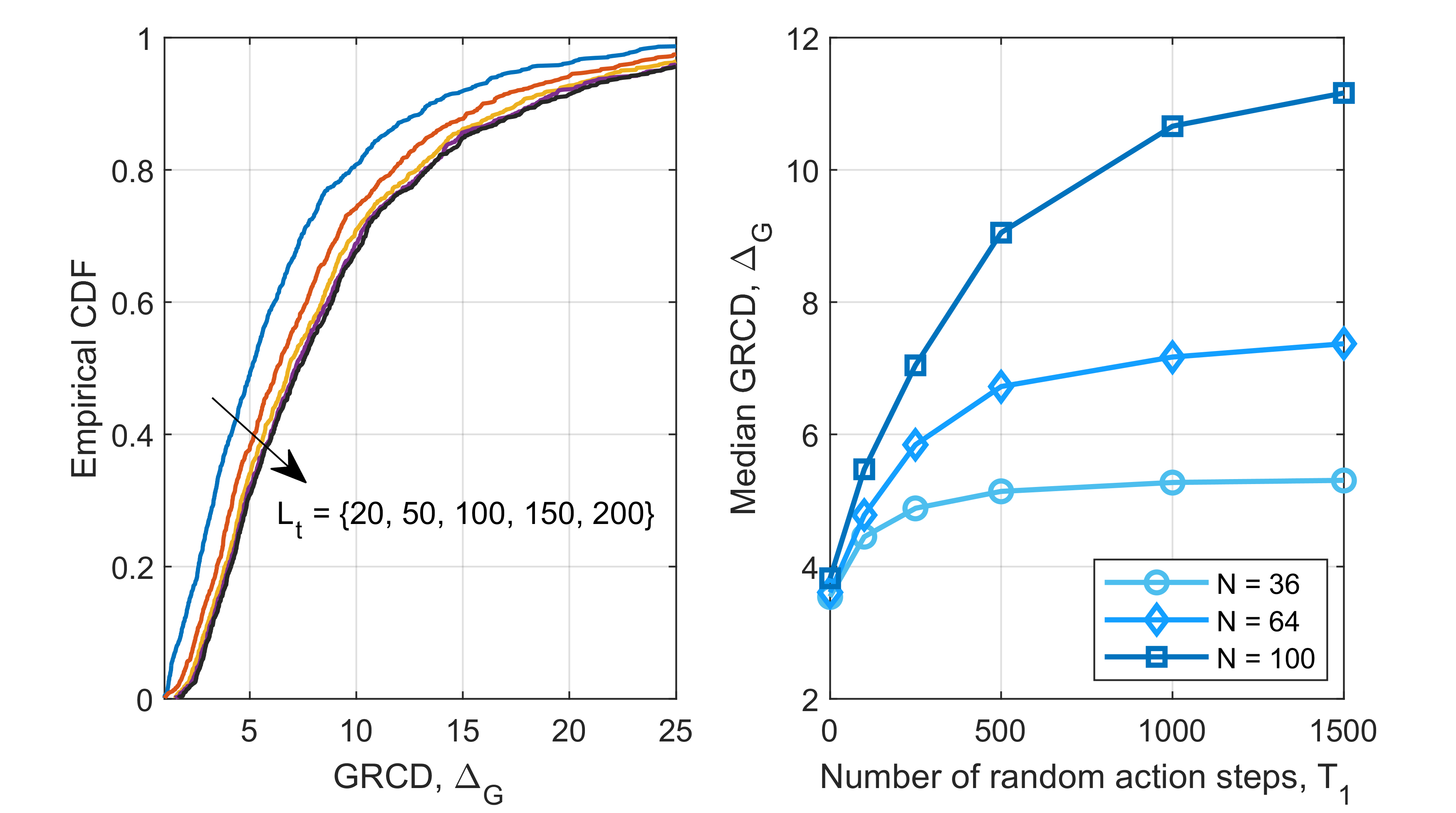}
\caption{Effect of (a) $L_{t}$ and (b) $T_{1}$ on the GRCD.}
\label{fig:4}
\end{figure}

Fig.~\ref{fig:4}(a) shows the effect of the training symbol duration $L_{t}$ on the quality of the solution obtained using Algorithm 1, where $N = 64$. One may observe that the GRCD obtained under small $L_{t}$ is fairly poor, due to the inaccurate estimation of the channel covariance matrices based on few signal samples against a large number of channels. Thus, longer symbols are required for accurate estimation. Reasonable performance is achieved once $L_{t}$ becomes moderately large (e.g., $L_{t} = 100$), beyond which diminishing returns are observed. These results suggest that a fairly accurate covariance estimation may be achieved when $L_{t}$ reaches a certain level compared to the total number of implicitly observed channels. It should be noted that long training symbols are needed only in the training phase. Once a satisfactory set of $\{\mathbf{g}, \mathbf{\Theta}\}$ is obtained, the data symbol duration can be reduced without affecting the GRCD.

Fig.~\ref{fig:4}(b) shows the convergence quality of Algorithm 1, in terms of the effect of the random training phase length on the median GRCD. We find that the GRCD increases with longer training phases to a certain extent; while having $T_{1} = 0$ results in very poor GRCD performance. As such, a random training phase is critical to mitigating the no-CSI nature of the problem. Note that for small $N$, $T_{1}$ may be reduced to shorten the training phase without much BER penalty.


\vspace{-2mm}
\section{Conclusion}
In this letter, under unknown CSI and ambient signal, we proposed a DRL-based framework to optimize the design parameters of an IRS-assisted AmBC system based on actual signal observations. The lack of CSI and varying reward function in each episode were mitigated with independent training in each episode and zero discount factor. Moreover, we utilized the optimal eigenvector combiner without IRS as initialization in each DRL step for effective exploration. Our results indicated that the DRL framework is capable of similar performance compared to various full-CSI benchmarks. The design of the DRL framework to take advantage of time-correlated channels is a useful avenue for further work.

\vspace{-2.5mm}
\bibliographystyle{ieeetran}
\bibliography{IEEEabrv,irs_ambc_ref}

\end{document}